# Alternative N=2 Supergravity in Five Dimensions with Singularities[1]

Hitoshi NISHINO[2]

*Department of Physics*
*University of Maryland*
*College Park, MD 20742-4111*

and

Subhash RAJPOOT[3]

*Department of Physics & Astronomy*
*California State University*
*Long Beach, CA 90840*

## Abstract

We present an alternative $N=2$ supergravity multiplet coupled to $n$ copies of vector multiplets and $n'$ copies of hypermultiplets in five dimensions. Our supergravity multiplet contains a single antisymmetric tensor and a dilaton, which are natural Neveu-Schwarz massless fields in superstring theory. The absence of the explicit Chern-Simons terms in our lagrangian deletes the non-trivial constraints on the couplings of vector multiplets in the conventional formulation. The scalars in the vector multiplets form the $\sigma$-model for the coset $SO(n,1)/SO(n)$, like those in the vector multiplets coupled to $N=1$ supergravity in nine dimensions, while the scalars in the hypermultiplets form that for the quaternionic Kähler manifold $Sp(n',1)/Sp(n')\times Sp(1)$. We also perform the gauging of the $SO(2)$ subgroup of the $Sp(1) = SL(2,\mathbb{R})$ automorphism group of $N=2$ supersymmetry. Our result is also generalized to singular 5D space-time as in the conventional formulation, as a preliminary for supersymmetric Randall-Sundrum brane world scenario.

PACS: 04.50.+h, 04.65.+e, 11.30.Pb
Key Words: Supergravity, Kaluza-Klein, Five-Dimensions

---

[1]This work is supported in part by NSF grant # PHY-98-02551.
[2]E-Mail: nishino@nscpmail.physics.umd.edu
[3]E-Mail: rajpoot@csulb.edu

## 1. Introduction

The importance of five-dimensional (5D) supergravity has revealed in many contexts of superstring [1] or M-theory [2], such as holographic Anti-de Sitter/superconformal field theory (AdS/SCF) correspondence, *i.e.*, a conjecture that the large $N$ limit of $SU(N)$ superconformal field theories in $D$ dimensions are dual equivalent to supergravity on AdS in $D+1$ dimensions [3][4]. The importance of studying 5D supergravity in AdS space-time is also motivated by the Randall-Sundrum scenario [5] for getting a large mass hierarchy by adjusting the tension of the 3-branes (domain walls) in 5D AdS [6][7][8].

The conventional on-shell formulation [9] of $N=1$ supergravity in 5D used in these studies was initiated in [10], which an arbitrary number of vector multiplets are coupled to supergravity, and generalized further in [11][12]. However, the drawback of these formulations [10][11][12] is the complication of couplings of an antisymmetric tensor $B_{\mu\nu}$ and a dilaton field to supergravity, which are important Neveu-Schwarz (NS) fields for superstring theory [1]. For example, the tensor fields in [10][11][12] always appear in symplectic pairs, due to their 'self-duality' condition in odd space-time dimensions [13], but the single antisymmetric tensor field $B_{\mu\nu}$ as the natural NS field [1] is not included in [11][12]. As an alternative approach, we may try off-shell formulations [14], but the drawback here is the lack of manifest $\sigma$-model geometrical structure formed by scalars inherent in the vector couplings to supergravity, which is 'hidden' at the off-shell level before eliminating auxiliary fields. This is similar to the 4D case of Kähler manifold structure in on-shell $N=1$ supergravity, which is hidden in the off-shell formulation [15].

In our present paper, we try to simplify vector multiplet couplings to supergravity, proposing an alternative on-shell $N=2$ supergravity multiplet in 5D, which has an on-shell irreducible field content larger than the conventional one [10][11][12], including the important antisymmetric tensor and dilaton fields. Our field content of supergravity multiplet is $(e_\mu{}^m, \psi_\mu{}^A, B_{\mu\nu}, \chi^A, A_\mu, \sigma)$ with 12+12 on-shell degrees of freedom, where the fünfbein $e_\mu{}^m$, the gravitini $\psi_\mu{}^A$, and the graviphoton $A_\mu$ coincide with the conventional $N=2$ supergravity [10][11][12], while an antisymmetric tensor $B_{\mu\nu}$, a dilatino $\chi^A$, and a dilaton $\sigma$ are our new field content. The important ingredient here is that the antisymmetric tensor $B_{\mu\nu}$ and the dilaton $\sigma$ are natural NS massless fields expected from superstring theory [1]. We also couple $n$ copies of vector multiplets $(C_\mu, \lambda^A, \varphi)$ to this $N=2$ supergravity, where the $n$ copies of the scalar $\varphi$ form the coordinates of the $\sigma$-model for the coset $SO(n,1)/SO(n)$. This coupling structure is similar to that for vector multiplets coupled to $N=1$ supergravity in 9D [16], and we have *no* explicit Chern-Simons term in the lagrangian. We further show how to gauge the $SO(2)$ subgroup of the $Sp(1) = SL(2,\mathbb{R})$ automorphism group of $N=2$ supersymmetries with the minimal couplings of these $n+1$ vector fields in a way similar to the gaugings in $N=1$ supergravity in 9D [17].

## 2. Alternative $N=2$ Supergravity Multiplet in 5D

We first clarify the structure of our alternative $N=2$ supergravity multiplet in 5D, which is distinct from the conventional one [10][11][12], as the foundation of any future elaboration of matter couplings. The field content of our supergravity multiplet is $(e_\mu{}^m, \psi_\mu{}^A, A_\mu, B_{\mu\nu}, \chi^A, \sigma)$ with $12+12$ on-shell degrees of freedom, where the first three fields coincide with the conventional supergravity multiplet. Here the indices $\mu, \nu, \cdots$ are for the curved world indices, $m, n, \cdots$ are local Lorentz with the signature $(\eta_{mn}) = \text{diag.} (-,+,+,+,+)$, while $A, B, \cdots = 1, 2$ are for the **2**-representation of the automorphism group $Sp(1) = SL(2,\mathbb{R})$ for the $N=2$ supersymmetry. In this paper, we use $Sp(1) = SL(2,\mathbb{R})$ notation instead of $SU(2)$ as the automorphism group, in order to make all the bosonic fields manifestly real, just for simplicity. The last three fields $B_{\mu\nu}, \chi^A, \sigma$ differentiate our multiplet from the conventional one [10].

Some new feature is elucidated by an invariant lagrangian for our supergravity multiplet:

$$\begin{aligned}\mathcal{L}_{\text{SG}} =\ & -\tfrac{1}{4}R - \tfrac{1}{2}(\overline{\psi}_\mu{}^A \gamma^{\mu\nu\rho} D_\nu \psi_{\rho A}) - \tfrac{1}{4}e^{-2\sigma} F^2_{\mu\nu} - \tfrac{1}{12}e^{-4\sigma} G^2_{\mu\nu\rho} - \tfrac{1}{2}(\overline{\chi}^A \gamma^\mu D_\mu \chi_A) - \tfrac{3}{4}(\partial_\mu \sigma)^2 \\ & - \tfrac{1}{24}e^{-2\sigma}(\overline{\psi}_\mu{}^A \gamma^{\mu\nu\rho\sigma\tau} \psi_{\nu A}) G_{\rho\sigma\tau} + \tfrac{1}{4}e^{-2\sigma}(\overline{\psi}_\mu{}^A \gamma_\nu \psi_{\rho A}) G^{\mu\nu\rho} \\ & - \tfrac{i}{2\sqrt{2}}e^{-\sigma}(\overline{\psi}_\mu{}^A \psi_{\nu A}) F^{\mu\nu} - \tfrac{i}{4\sqrt{2}}e^{-\sigma}(\overline{\psi}_\mu{}^A \gamma^{\mu\nu\rho\sigma} \psi_{\nu A}) F_{\rho\sigma} \\ & - \tfrac{1}{2\sqrt{6}}e^{-\sigma}(\overline{\psi}_\mu{}^A \gamma^{\rho\sigma} \gamma^\mu \chi_A) F_{\rho\sigma} + \tfrac{i}{6\sqrt{3}}e^{-2\sigma}(\overline{\psi}_\mu{}^A \gamma^{\rho\sigma\tau} \gamma^\mu \chi_A) G_{\rho\sigma\tau} \\ & - \tfrac{5}{72}e^{-2\sigma}(\overline{\chi}^A \gamma^{\mu\nu\rho} \chi_A) G_{\mu\nu\rho} - \tfrac{i}{12\sqrt{2}}e^{-\sigma}(\overline{\chi}^A \gamma^{\mu\nu} \chi_A) F_{\mu\nu} + \tfrac{\sqrt{3}i}{2}(\overline{\psi}_\mu{}^A \gamma^\nu \gamma^\mu \chi_A) \partial_\nu \sigma \ , \end{aligned} \quad (2.1)$$

up to a total divergence and quartic fermion terms, under supersymmetry

$$\begin{aligned}\delta_Q e_\mu{}^m &= +(\overline{\epsilon}^A \gamma^m \psi_{\mu A}) \ , \qquad \delta_Q \sigma = +\tfrac{i}{\sqrt{3}}(\overline{\epsilon}^A \chi_A) \ , \\ \delta_Q \psi_\mu{}^A &= +D_\mu(\widehat{\omega})\epsilon^A + \tfrac{i}{6\sqrt{2}}e^{-\sigma}(\gamma_\mu{}^{\rho\sigma} - 4\delta_\mu{}^\rho \gamma^\sigma)\epsilon^A F_{\rho\sigma} + \tfrac{1}{18}e^{-2\sigma}(\gamma_\mu{}^{\rho\sigma\tau} - \tfrac{3}{2}\delta_\mu{}^\rho \gamma^{\sigma\tau})\epsilon^A G_{\rho\sigma\tau} \ , \\ \delta_Q A_\mu &= -\tfrac{i}{\sqrt{2}}e^\sigma(\overline{\epsilon}^A \psi_{\mu A}) + \tfrac{1}{\sqrt{6}}e^\sigma(\overline{\epsilon}^A \gamma_\mu \chi_A) \ , \\ \delta_Q B_{\mu\nu} &= +e^{2\sigma}(\overline{\epsilon}^A \gamma_{[\mu} \psi_{\nu]A}) + \tfrac{i}{\sqrt{3}}e^{2\sigma}(\overline{\epsilon}^A \gamma_{\mu\nu} \chi_A) + 2A_{[\mu|}\delta_Q A_{|\nu]} \ , \\ \delta_Q \chi^A &= -\tfrac{1}{2\sqrt{6}}e^{-\sigma}\gamma^{\mu\nu}\epsilon^A F_{\mu\nu} + \tfrac{i}{6\sqrt{3}}e^{-2\sigma}\gamma^{\mu\nu\rho}\epsilon^A G_{\mu\nu\rho} - \tfrac{\sqrt{3}i}{2}\gamma^\mu \epsilon^A \partial_\mu \sigma \ , \end{aligned} \quad (2.2)$$

up to quadratic fermion terms. As usual with antisymmetric tensors [9], $G_{\mu\nu\rho}$ is the field strength of $B_{\mu\nu}$ modified by a Chern-Simons form:

$$G_{\mu\nu\rho} \equiv 3\partial_{[\mu} B_{\nu\rho]} + 3F_{[\mu\nu} A_{\rho]} \ . \quad (2.3)$$

The closure of supersymmetries can be easily confirmed with the parameter of translation

$$\xi^m \equiv +i(\overline{\epsilon}_2{}^A \gamma^m \epsilon_{1A}) \equiv +i(\overline{\epsilon}_2 \gamma^m \epsilon_1) \ . \quad (2.4)$$

As in the second expression here, we omit from now on the explicit contracted indices $A, B, \cdots$.

As in the usual dilaton couplings in supergravity [9], the antisymmetric field $B_{\mu\nu}$ and the graviphoton $A_\mu$ are scaled, when the dilaton $\sigma$ is shifted by a constant value:

$$\sigma \to \sigma + c \ , \qquad B_{\mu\nu} \to e^{2c} B_{\mu\nu} \ , \qquad A_\mu \to e^c A_\mu \ , \quad (2.5)$$

where $c$ is an arbitrary constant parameter. This global symmetry controls the various exponential couplings of $\sigma$ in the lagrangian (2.1).

The derivations of the transformation rule (2.2) and the lagrangian (2.1) are rather routine, described as follows. We start with putting unknown coefficients for terms in (2.2), which are determined by the closure of supersymmetry at the linear level on all the bosonic fields, and the invariance of the kinetic terms in (2.1). These two sets of conditions fix all the unknown coefficients up to the sign ambiguities for field redefinitions. This fixes all the terms in (2.2) up to quadratic fermion terms. Now for the remaining Noether and Pauli couplings in (2.1), we put unknown coefficients, in addition to the coefficient for the Chern-Simons term in (2.3), which are determined by the cancellations of (fermion)×(boson)$^2$-type terms after the variation of such lagrangian under the already-fixed rule (2.2). The structures of these terms are (i) $\psi G^2$, (ii) $\psi F^2$, (iii) $\psi(\partial\sigma)^2$, (iv) $\psi FG$, (v) $\psi F\partial\sigma$, (vi) $\psi G\partial\sigma$, (vii) $\chi G^2$, (viii) $\chi F^2$, (ix) $\chi(\partial\sigma)^2$, (x) $\chi FG$, (xi) $\chi F\partial\sigma$, and (xii) $\chi G\partial\sigma$. All of these sectors consistently fix the coefficients as in (2.1), only up to field redefinitions.

Compared with the conventional formulations [10], there is a similarity as well as basic difference, which will be more manifest after coupling to vector multiplets in the next section. The similarity is that our tensor field $B_{\mu\nu}$ can be dualized into a vector field $B_\mu$ by a duality transformation [18], so that the final field content will be $(e_\mu{}^m, \psi_\mu{}^A, A_\mu, B_\mu, \chi^A, \sigma)$. ¿From this viewpoint, our system (2.1) is 'dual equivalent' to the conventional formulation with only one vector multiplet, in particular the dilaton field plays the coordinate of $SO(1,1)$, as usual in superstring theory. However, the caveat at this stage is that even though such a duality transformation is possible even after coupling vector multiplets, the resulting $\sigma$-model structure is qualitatively different from that given in the conventional formulations [10][11][12], as will be seen shortly.

Some readers may be still wondering how a supermultiplet with spins ranging from two to zero (in the 4D sense) can be an 'irreducible' multiplet. There is nothing puzzling about this from the viewpoint that in higher dimensions in $D > 4$, the usual irreducibility from 4D viewpoint does not hold. A typical supporting example for this statement is the $N = 2$ supergravity multiplet in 6D [19]. When we require the existence of an action, the smallest supermultiplet with sexbein has the field content $(e_\mu{}^m, \psi_\mu{}^A, B_{\mu\nu}, \chi^A, \varphi)$, similarly to our supermultiplet above in 5D. Here $N = 2$ supergravity in 6D yields maximally $N = 2$ supergravity in 4D after a simple dimensional reduction. So naïvely from 4D viewpoint, we expect the 'irreducible' supergravity supermultiplet in 6D only with spins 2, 3/2 and 1, and therefore it appears to be strange to have even a scalar dilaton $\varphi$ here. However, it is well-known that this is the 'smallest' supermultiplet in 6D, as long as an invariant action is required [19].[4] We repeat here that the 'irreducibility' in the 4D sense works no longer in higher dimensions, when supermultiplets in 4D are 'unified' in $D > 4$.

Notice that the antisymmetric field $B_{\mu\nu}$ and the dilaton $\sigma$ are the natural NS massless fields in superstring [1] or M-theory [2]. In other words, it is more natural to have a supergravity with these fields in the point field theory limit. Another advantage of introducing an antisymmetric tensor $B_{\mu\nu}$ is associated with the recent development of non-commutative geometry [20] in which the tensor $B_{\mu\nu}$ develops certain non-trivial constant value. We stress the fact that our supergravity multiplet contains the NS fields $B_{\mu\nu}$ and $\sigma$ as irreducible component fields, indicating that our supergravity is a more natural point field theory limit of superstring theory [1] than the conventional one [10][11][12].

---

[4]Here we do not go into the subtlety about the existence of action principle in this context, but the significance implied in this context is clear.

## 3. Couplings to Vector Multiplets and Hypermultiplets

Our next task is to couple our $N=2$ supergravity to $n$ copies of vector multiplets and $n'$ copies of hypermultiplets, as we have promised. The field content of a vector multiplet is $(B_\mu, \chi^A, \varphi)$ with $4+4$ degrees of freedom, and therefore we expect that the coupling structure must be parallel to the case in 9D [16][17], in which the scalars form the coordinates of the $\sigma$-model on the coset $H^n \equiv SO(n,1)/SO(n)$ [16], as the simplest case of symmetric non-Jordan family scalar manifold [10]. As for the hypermultiplets with the field content $(\phi^{\underline{\alpha}}, \psi^{\underline{a}})$ with $4n' + 4n'$ degrees of freedom, the couplings are in such a way that the scalars $\phi^{\underline{\alpha}}$ form the $4n'$-dimensional coordinates of quaternionic Kähler manifold $HP(n'-1,1) \equiv Sp(n',1)/Sp(n') \times Sp(1)$, like those in $N=2$ supersymmetry in 4D [21] or in 6D [19].

For readers who are not very familiar with the general backgrounds of determining a coset in a locally supersymmetric $\sigma$-model, we give a brief remark here. The above coset $H^n \equiv SO(n,1)/SO(n)$ for the vector multiplets was determined as the most natural candidate, due to the similarity of our $N=2$ supergravity multiplet to $N=1$ supergravity in 9D with similar ('formally' the same) field content $(e_\mu{}^m, \psi_\mu, B_{\mu\nu}, A_\mu, \chi, \varphi)$ [16][17]. The reason of this similarity is due to the repeating structure of Clifford algebra in general higher dimensions [22], i.e., some aspects of supersymmetry repeat themselves every four space-time dimensions. Furthermore, the field content of an $N=2$ vector multiplet in 5D is also similar to $N=1$ vector multiplet in 9D, in which a scalar field belongs to a vector multiplet. Accordingly, when $n$ copies of vector multiplets are coupled to supergravity, there arise in total $n+1$ scalar fields: one from the supergravity multiplet (dilaton), and $n$ copies from the $n$ vector multiplets. In such a case, we need some $\sigma$-model structure in order to realize local supersymmetry. This is because we need some type of 'complex structure' in supersymmetry transformation rules for scalar fields, as is also the case with $N=2$ local supersymmetry in 4D [21]. In 9D [16][17], it was found that when $n$ vector multiplets are coupled to supergravity, the $n$ copies of scalar fields in these vector multiplets form the $n$-dimensional coordinates with the positive definite signature, while the dilaton in the supergravity multiplet form a special coordinate with the negative signature, thus forming the coset $H^n \equiv SO(n,1)/SO(n)$. The indefinite signature of the coset due to the two sets of scalar fields is related to the difference between matter versus supergravity fields, similar to the difference between the dS and AdS minima of potentials, respectively. Even though, in principle, we can generalize this coset to more sophisticated non-Jordan family scalar cosets as in [10], we focus on this coset $H^n$ just for simplicity in this paper.

In order to clarify certain important geometry of $H^n \equiv SO(n,1)/SO(n)$, we start with the coset algebra for the coset generators $K_a$, and $SO(n)$ generators $H_{ab}$ [23]:

$$[H_{ab}, H_{cd}] = \delta_{bc} H_{ad} - \delta_{ac} H_{bd} + \delta_{ad} H_{bc} - \delta_{bd} H_{ac} ,$$

$$[H_{ab}, K_c] = \delta_{bc} K_a - \delta_{ac} K_b , \quad [K_a, K_b] = +\tfrac{1}{\xi^2} H_{ab} . \qquad (3.1)$$

Here the indices $a, b, \cdots = (1), (2), \cdots, (n)$ are for the vectorial representation of $SO(n)$. The numerical constant $\xi$ in the last line is *a priori* undetermined by the geometry alone, but will be determined by the action invariance under supersymmetry. The coset representatives $L_I{}^A$ satisfy the relationship

$$L_A{}^I \partial_\alpha L_I{}^B = \tfrac{1}{2} A_\alpha{}^{ab} (H_{ab})_A{}^B + V_\alpha{}^a (K_a)_A{}^B . \qquad (3.2)$$

The indices $A, B, \cdots = ((0),a), ((0),b), \cdots = (0), (1), (2), \cdots, (n)$ are for the local coordinates on $H^n$.[5] In other words, $A, B, \cdots = ((0),a), ((0),b), \cdots$ are the $(n+1)$-dimensional extension of the original $n$-dimensional indices $a, b, \cdots$. The indices $I, J, \cdots = 0, 1, \cdots, n$ are for the curved coordinates, while $\alpha, \beta, \cdots = 1, 2, \cdots, n$ are for the coordinates on $H^n$. The raising/lowering of the indices $A, B, \cdots$ is done by the metric tensor $(\eta_{AB}) = \text{diag.}(-,+,+,\cdots,+)$. In (3.2), the generators $H$'s and $K$'s have the components

$$(H_{ab})_c{}^d = \delta_{ac}\delta_b{}^d - \delta_{bc}\delta_a{}^d \ , \qquad (K_a)_{b(0)} = (K_a)_{(0)b} = \tfrac{1}{\xi}\delta_{ab} \ , \tag{3.3}$$

which are the only non-zero components of these generators. The $V_\alpha{}^a$ are the vielbein for $H^n$, while $A_\alpha{}^{ab}$ is the $SO(n)$ composite connection, which acts like

$$D_\alpha X_a \equiv \partial_\alpha X_a + A_{\alpha a}{}^b X_b \ , \tag{3.4}$$

for an arbitrary $SO(n)$ vector $X_a$. Due to the ortho-normality

$$L_I{}^A L_A{}^J = \delta_I{}^J \ , \qquad L_A{}^I L_I{}^B = \delta_A{}^B \ , \tag{3.5}$$

it is convenient to define

$$L_I \equiv L_I{}^{(0)} \ , \qquad L^I \equiv L_{(0)}{}^I \ , \tag{3.6}$$

satisfying

$$L_I L^I = +1 \ , \qquad L_a{}^I L_I \equiv 0 \ , \qquad L_I{}^a L^I \equiv 0 \ . \tag{3.7}$$

¿From the Maurer-Cartan form (3.2), it follows that

$$D_\alpha L_I = \partial_\alpha L_I = \tfrac{1}{\xi} L_I{}^a V_{\alpha a} \ , \qquad D_\alpha L_I{}^a = \tfrac{1}{\xi} L_I V_\alpha{}^a \ , \tag{3.8}$$

so that the tensor $L_{IJ}$ defined by

$$L_{IJ} \equiv \eta_{AB} L_I{}^A L_J{}^B = -L_I L_J + L_I{}^a L_{Ja} \tag{3.9}$$

satisfies the 'constancy' condition of $L_{IJ}$:

$$\partial_\alpha L_{IJ} = 0 \ . \tag{3.10}$$

Therefore we can choose the frame such that this $(n+1) \times (n+1)$ matrix is diagonal: $(L_{IJ}) = \text{diag.}(-,+,+,\cdots,+)$. Other relevant useful relations are such as

$$L_{IJ} L^J = -L_I \ , \qquad L_{IJ} L_a{}^J = +L_{Ia} \ . \tag{3.11}$$

We can also get the commutator

$$[D_\alpha, D_\beta] L_I{}^a = -\tfrac{1}{\xi^2}(V_\alpha{}^a V_\beta{}^b - V_\beta{}^a V_\alpha{}^b) L_{Ib} \ , \tag{3.12}$$

---

[5]Even though we are using the same indices $A, B, \cdots$ both for the **2**-representations and for these local Lorentz coordinates, they are not confusing, as long as we keep track of the context they are used.

leading to the curvature tensor of the manifold $H^n$

$$R_{\alpha\beta}{}^{ab} = -\tfrac{1}{\xi^2}(V_\alpha{}^a V_\beta{}^b - V_\beta{}^a V_\alpha{}^b) \ , \tag{3.13}$$

with the negative definite constant scalar curvature $R = -n(n-1)/\lambda^2 \leq 0$. As will be seen, the value of $\lambda$ in the above relationships will be determined to be

$$\xi = -\tfrac{1}{\sqrt{2}} \ , \tag{3.14}$$

by the action invariance under supersymmetry.

We can perform the coupling of these vector multiplets to our supergravity (2.1), following the previous results of the $N=1$ supergravity in 9D [16][17] for vector multiplet couplings. For example, we see the right assignment for the gaugino $\lambda$ to be the **n**-representation under $SO(n)$, while the **2**-representation under the $Sp(1) = SL(2,\mathbb{R})$ group. Relevantly, the graviphoton $A_\mu$ in the original supergravity multiplet (2.2) is to be identified as the 0-th component of $A_\mu{}^I$.

As for the coupling of the hypermultiplets $(\phi^{\underline{\alpha}}, \psi^{\underline{a}})$ to supergravity, the scalars $\phi^{\underline{\alpha}}$ ($\underline{\alpha} = 1, 2, \cdots, 4n'$) form $4n'$-dimensional coordinates of a quaternionic Kähler manifold $HP(n'-1,1) \equiv Sp(n',1)/Sp(n') \times Sp(1)$, while $\psi^{\underline{a}}$ ($\underline{a} = 1, \cdots, 2n'$) transform as $\mathbf{2n'}$-representation of $Sp(n')$, as in 4D [21] or in 6D [19]. As in the case of vector multiplets, we need here similar geometrical preliminary for the manifold $HP(n'-1,1)$ [21][19]. We start with the representative $L^{\underline{a}\underline{\alpha}}$, which satisfy the Maurer-Cartan form for the coset $HP(n'-1,1)$:

$$L^{-1}\partial_{\underline{\alpha}} L = A_{\underline{\alpha}}{}^i T_i + A_{\underline{\alpha}}{}^{\underline{I}} T_{\underline{I}} + V_{\underline{\alpha}}{}^{\underline{a}A} K_{\underline{a}A} \ . \tag{3.15}$$

The $(T_{\underline{I}})_{\underline{a}}{}^{\underline{b}}$ ($\underline{I}, \underline{J}, \cdots = 1, 2, \cdots, n'(2n'+1)$) are the generators of $Sp(n')$, and $(T^i)_A{}^B$ ($i, j, \cdots = 1, 2, 3$) are the generators of the automorphism group $Sp(1) = SL(2,\mathbb{R})$, while $K_{\underline{a}A}$ are the generators in the coset space. The indices $A, B, \cdots = 1, 2$ for the automorphism group $Sp(1)$ are the same as before. The $Sp(1)$ generators $T_i$ are anti-hermitian, and have symmetric components: $(T_i)_{AB} = +(T_i)_{BA}$, $(T_i)_A{}^B (T_j)_B{}^C = -(1/4)\delta_{ij}\delta_A{}^C + (1/2)\epsilon_{ij}{}^k (T_k)_A{}^C$, and $(T_i)_A{}^B \equiv \epsilon^{BC}(T_i)_{AC}$ with the $Sp(1) = SL(2,\mathbb{R})$ metric $\epsilon^{AB} = -\epsilon^{BA}$, and *idem for* $Sp(n')$. These notations are the same as that in [19], except for the *underlined* indices to be distinguished from those in the previous coset $H^n \equiv SO(n,1)/SO(n)$.

Skipping other details, we give the ortho-normality conditions of vielbeins:[6]

$$g_{\underline{\alpha}\underline{\beta}} V_{\underline{a}A}{}^{\underline{\alpha}} V_{\underline{b}B}{}^{\underline{\beta}} = \epsilon_{\underline{ab}} \epsilon_{AB} \ , \qquad 2 V_{\underline{a}A}{}^{(\underline{\alpha}|} V^{\underline{a}B|\underline{\beta})} = g^{\underline{\alpha}\underline{\beta}} \delta_A{}^B \ , \tag{3.16}$$

with the metric on $HP(n'-1,1)$ and the antisymmetric invariant $\epsilon_{\underline{ab}}$ of $Sp(n')$. Since $HP(n'-1,1)$ is a quaternionic Kähler manifold, it has a triple of covariantly constant complex structures $J_{\underline{\alpha}\underline{\beta}}{}^i$ defined by [21][19]

$$J_{\underline{\alpha}\underline{\beta}}{}^i = -(T^i)_A{}^B (V_{\underline{\alpha}\underline{a}B} V_{\underline{\beta}}{}^{\underline{a}A} - V_{\underline{\beta}\underline{a}B} V_{\underline{\alpha}}{}^{\underline{a}A}) \ , \tag{3.17}$$

---

[6] Notice that there was a similar equation $2V_{\underline{a}A}{}^{(\underline{\alpha}|} V_{\underline{b}}{}^{A|\underline{\beta})} = n'^{-1} g^{\underline{\alpha}\underline{\beta}} \epsilon_{\underline{ab}}$ originally in [21]. However, we note that this equation is *not* correct, as can be easily seen by multiplying this by $V_{\underline{\alpha}}{}^{\underline{a}B}$, yielding $2n' - 1 = n'^{-1}$ which holds only for $n' = 1$. There must be an additional term antisymmetric in $\underline{a} \leftrightarrow \underline{b}$ on the r.h.s. of the above equation. Fortunately, this equation has not been used in the supersymmetry invariance in ordinary supergravity formulations [21][19].

satisfying the quaternion algebra $J^i J^j = -\delta^{ij} + \epsilon^{ijk} J^k$. As in [21][19], the composite $Sp(1)$ connection $A_{\underline{\alpha}}{}^i$ couples to the gravitino $\psi_\mu{}^A$, while the composite $Sp(n')$ connection $A_{\underline{\alpha}}{}^I$ couples to the fermions $\psi^{\underline{a}}$. Due to the quaternionic geometry, these quaternions and the $Sp(1)$ curvature $F_{\underline{\alpha}\underline{\beta}}{}^i$ are proportional to each other:

$$F_{\underline{\alpha}\underline{\beta}}{}^i = \eta J_{\underline{\alpha}\underline{\beta}}{}^i = -2 J_{\underline{\alpha}\underline{\beta}}{}^i \quad , \tag{3.18}$$

with an *a priori* unknown numerical constant $\eta$, which will be fixed to be $\eta = -2$ by the invariance of the total action.

Armed with these geometrical relationships at hand, it is now straightforward to couple the vector multiplets and hypermultiplets to supergravity. After all, our total field content is such that our supergravity multiplet $(e_\mu{}^m, \psi_\mu{}^A, B_{\mu\nu}, \chi^A, A_\mu, \sigma)$ and the $n$ copies of the vector multiplets $(B_\mu, \lambda^A, \varphi)$, as well as the $n'$ copies of the hypermultiplets $(\phi^{\underline{\alpha}}, \psi^{\underline{a}})$ are 'fused' together to yield $(e_\mu{}^m, \psi_\mu{}^A, B_{\mu\nu}, \chi^A, A_\mu{}^I, \lambda^{aA}, \varphi^\alpha, \sigma, \phi^{\underline{\alpha}}, \psi^{\underline{a}})$, where $\varphi^\alpha$ are for the $\sigma$-model coordinates of $H^n$, and $\phi^{\underline{\alpha}}$ are for those of $HP(n'-1,1)$. Now our invariant lagrangian (up to a total divergence and quartic fermion terms) thus obtained is

$$\begin{aligned}\mathcal{L}_0 = &-\tfrac{1}{4}R - \tfrac{1}{2}(\overline{\psi}_\mu \gamma^{\mu\nu\rho} D_\nu \psi_\rho) - \tfrac{1}{12} e^{-4\sigma} G^2_{\mu\nu\rho} - \tfrac{1}{4} e^{-2\sigma}(L_I{}^a L_{Ja} + L_I L_J) F_{\mu\nu}{}^I F^{\mu\nu\,J} \\ &- \tfrac{1}{2}(\overline{\lambda}^a \gamma^\mu D_\mu \lambda_a) - \tfrac{1}{2} g_{\alpha\beta}(\partial_\mu \varphi^\alpha)(\partial^\mu \varphi^\beta) - \tfrac{3}{4}(\partial_\mu \sigma)^2 - \tfrac{1}{2}(\overline{\chi}\gamma^\mu D_\mu \chi) \\ &- \tfrac{1}{2} g_{\underline{\alpha}\underline{\beta}}(\partial_\mu \phi^{\underline{\alpha}})(\partial^\mu \phi^{\underline{\beta}}) - \tfrac{1}{2}(\overline{\psi}^{\underline{a}} \gamma^\mu \partial_\mu \psi_{\underline{a}}) \\ &+ \tfrac{i}{\sqrt{2}} V_\alpha{}^a (\overline{\psi}_\mu \gamma^\nu \gamma^\mu \lambda^a) \partial_\nu \varphi^\alpha + \tfrac{\sqrt{3}i}{2}(\overline{\psi}_\mu \gamma^\nu \gamma^\mu \chi) \partial_\nu \sigma + i V_{\underline{\alpha}}{}^{aA}(\overline{\psi}_\mu \gamma^\nu \gamma^\mu \psi_{\underline{a}}) \partial_\nu \phi^{\underline{\alpha}} \\ &- \tfrac{i}{4\sqrt{2}} e^{-\sigma}(\overline{\psi}_\mu \gamma^{\mu\nu\rho\sigma} \psi_\nu + 2\overline{\psi}^\rho \psi^\sigma) L_I F_{\rho\sigma}{}^I + \tfrac{i}{6\sqrt{3}} e^{-2\sigma}(\overline{\psi}_\mu \gamma^{\rho\sigma\tau} \gamma^\mu \chi) G_{\rho\sigma\tau} \\ &- \tfrac{1}{24} e^{-2\sigma}(\overline{\psi}_\mu \gamma^{\mu\rho\rho\sigma\tau} \psi_\nu - 6\overline{\psi}^\rho \gamma^\sigma \psi^\tau) G_{\rho\sigma\tau} - \tfrac{5}{72} e^{-2\sigma}(\overline{\chi}\gamma^{\mu\nu\rho}\chi) G_{\mu\nu\rho} \\ &- \tfrac{1}{2\sqrt{2}} e^{-\sigma}(\overline{\psi}_\mu \gamma^{\rho\sigma} \gamma^\mu \lambda_a) L_I{}^a F_{\rho\sigma}{}^I - \tfrac{1}{2\sqrt{6}} e^{-\sigma}(\overline{\psi}_\mu \gamma^{\rho\sigma}\gamma^\mu \chi) L_I F_{\rho\sigma}{}^I - \tfrac{i}{12\sqrt{2}} e^{-\sigma}(\overline{\chi}\gamma^{\mu\nu}\chi) L_I F_{\mu\nu}{}^I \\ &+ \tfrac{i}{4\sqrt{2}} e^{-\sigma}(\overline{\lambda}^a \gamma^{\mu\nu} \lambda_a) L_I F_{\mu\nu}{}^I + \tfrac{1}{24} e^{-2\sigma}(\overline{\lambda}^a \gamma^{\mu\nu\rho} \lambda_a) G_{\mu\nu\rho} + \tfrac{1}{24} e^{-2\sigma}(\overline{\psi}^{\underline{a}} \gamma^{\mu\nu\rho} \psi_{\underline{a}}) G_{\mu\nu\rho} \\ &+ \tfrac{i}{\sqrt{6}} e^{-2\sigma}(\overline{\chi}\gamma^{\mu\nu}\lambda_a) L_I{}^a F_{\mu\nu}{}^I - \tfrac{1}{4\sqrt{2}} e^{-\sigma}(\overline{\psi}^{\underline{a}} \gamma^{\mu\nu} \psi_{\underline{a}}) L_I F_{\mu\nu}{}^I \quad . \end{aligned} \tag{3.19}$$

Compared with the case with no vector multiplet (2.3), the field strength of $B_{\mu\nu}$ is now modified as

$$G_{\mu\nu\rho} \equiv 3\partial_{[\mu} B_{\nu\rho]} - 3 L_{IJ} F_{[\mu\nu}{}^I A_{\rho]}{}^J \quad . \tag{3.20}$$

The previous case of (2.3) is now a special case of (3.20), *i.e.*, only $L_{00} = -1$ is present. Similarly to the case of $N=1$ supergravity in 9D [16][17], the kinetic term for the vectors $A_\mu{}^I$ has a positive definite coefficient matrix in the combination $(L_{Ia} L_J{}^a + L_I L_J) = \text{diag.}(+, +, \cdots, +) + \mathcal{O}(\varphi)$, where the $\mathcal{O}(\varphi)$-terms yield just cubic or higher-order interactions. The covariant derivatives for $Sp(1)$ or $Sp(n')$-covariant spinors are

$$\begin{aligned} D_{[\mu|}\psi_{|\nu]}{}^A &\equiv D_{[\mu|}(\widehat{\omega})\psi_{|\nu]}{}^A + (\partial_{[\mu|}\phi^{\underline{\alpha}}) A_{\underline{\alpha}}{}^i (T_i \psi_{|\nu]})^A \quad , \\ D_\mu \chi^A &\equiv D_\mu(\widehat{\omega})\chi^A + (\partial_\mu \phi^{\underline{\alpha}}) A_{\underline{\alpha}}{}^i (T_i \chi)^A \quad , \end{aligned}$$

$$D_\mu \lambda^{aA} \equiv D_\mu(\widehat{\omega})\lambda^{aA} + (\partial_\mu \varphi^\alpha) A_{\underline{\alpha}}{}^{ab}\lambda_b{}^A + (\partial_\mu \phi^{\underline{\alpha}}) A_{\underline{\alpha}}{}^i (T_i \lambda^a)^A ,$$
$$D_\mu \psi^{\underline{a}} \equiv D_\mu(\widehat{\omega})\psi^{\underline{a}} + (\partial_\mu \phi^{\underline{\alpha}}) A_{\underline{\alpha}}{}^{\underline{I}}(T_{\underline{I}}\psi)^{\underline{a}} , \quad (3.21)$$

with the supercovariant Lorentz connection $\omega_\mu{}^{mn}$ [9]. The actions of the $Sp(1)$ and $Sp(n')$ generators $T_i$ and $T_{\underline{I}}$ are $e.g.$, $(T_i\epsilon)^A \equiv (T_i)^{AB}\epsilon_B$, $(T_{\underline{I}}\psi)^{\underline{a}} \equiv (T_{\underline{I}})^{\underline{ab}}\psi_{\underline{b}}$, etc. Our supersymmetry transformation rule is

$$\begin{aligned}
\delta_Q e_\mu{}^m &= +(\overline{\epsilon}\gamma^m \psi_\mu) , \quad \delta_Q \sigma = +\tfrac{i}{\sqrt{3}}(\overline{\epsilon}\chi) , \\
\delta_Q \psi_\mu{}^A &= +D_\mu \epsilon^A + \tfrac{i}{6\sqrt{2}} e^{-\sigma}(\gamma_\mu{}^{\rho\sigma} - 4\delta_\mu{}^\rho \gamma^\sigma)\epsilon^A L_I F_{\rho\sigma}{}^I + \tfrac{1}{18}e^{-2\sigma}(\gamma_\mu{}^{\rho\sigma\tau} - \tfrac{3}{2}\delta_\mu{}^\rho \gamma^{\sigma\tau})\epsilon^A G_{\rho\sigma\tau} , \\
\delta_Q A_\mu{}^I &= -\tfrac{i}{\sqrt{2}}e^\sigma(\overline{\epsilon}\psi_\mu) + \tfrac{1}{\sqrt{6}}e^\sigma L^I (\overline{\epsilon}\gamma_\mu \chi) + \tfrac{1}{\sqrt{2}}e^\sigma (\overline{\epsilon}\gamma_\mu \lambda^a) L_a{}^I , \\
\delta_Q B_{\mu\nu} &= +e^{2\sigma}(\overline{\epsilon}\gamma_{[\mu}\psi_{\nu]}) + \tfrac{i}{\sqrt{3}}e^{2\sigma}(\overline{\epsilon}\gamma_{\mu\nu}\chi) - 2L_{IJ} A_{[\mu|}{}^I \delta_Q A_{|\nu]}{}^J , \\
\delta_Q \chi^A &= -\tfrac{1}{2\sqrt{6}}e^{-\sigma}\gamma^{\mu\nu}\epsilon^A L_I F_{\mu\nu}{}^I + \tfrac{i}{6\sqrt{3}}e^{-2\sigma}\gamma^{\mu\nu\rho}\epsilon^A G_{\mu\nu\rho} - \tfrac{\sqrt{3}i}{2}\gamma^\mu \epsilon^A \partial_\mu \sigma , \\
\delta_Q \varphi^\alpha &= +\tfrac{i}{\sqrt{2}} V_a{}^\alpha(\overline{\epsilon}\lambda^a) , \\
\delta_Q \lambda^{aA} &= -\tfrac{1}{2\sqrt{2}} e^{-\sigma}\gamma^{\mu\nu}\epsilon^A L_I{}^a F_{\mu\nu}{}^I - \tfrac{i}{\sqrt{2}}\gamma^\mu \epsilon^A V_\alpha{}^a \partial_\mu \varphi^\alpha , \\
\delta_Q \phi^{\underline{\alpha}} &= +iV_{\underline{a}A}{}^{\underline{\alpha}}(\overline{\epsilon}^A \psi^{\underline{a}}) , \\
\delta_Q \psi^{\underline{a}} &= -iV_{\underline{\alpha}}{}^{\underline{a}A}\gamma^\mu \epsilon_A \partial_\mu \phi^{\underline{\alpha}} , \quad (3.22)
\end{aligned}$$

up to quadratic fermion terms.

Our derivations of (3.19) and (3.22) are outlined as follows. We first determine all the terms in (3.22) by putting unknown coefficients in each new terms in (3.22), other than those in (2.2). These coefficients are determined by the linear-level closure of supersymmetry on all the bosonic fields, and the invariance of the free kinetic terms for the vector multiplet, up to possible field redefinitions. Now using (3.22), we can fix all the Noether and Pauli couplings in (3.19) by a procedure similar to section 2, including also the $a$ $priori$ unknown constants $\xi$ in (3.14) and $\eta$ in (3.18). The value of $\xi$ is determined simultaneously by the following four sectors in the invariance confirmation of our lagrangian (3.19) under supersymmetry (3.22), which are all of the (fermion) $\times$ (boson)$^2$-type, categorized as (i) $\lambda F^2$-type from $(\delta_Q L_I) L_J F_{\mu\nu}{}^I F^{\mu\nu J}$, (ii) $\lambda F \partial\varphi$-type from $(\delta_Q \psi)\lambda LF$, (iii) $\chi F \partial \varphi$-type from $L_I L_J F_{\mu\nu}{}^I F^{\mu\nu J}$ or $\psi \chi L_I F$, and (iv) $\psi F \partial\varphi$-type from either $L_I L_J (\delta_Q F_{\mu\nu}{}^I) F^{\mu\nu J}$ or $(\delta_Q \psi)\psi LF$, where $F$, $\psi$, $\lambda$ and $L$ symbolize $\psi_\mu$, $\lambda^a$, $F_{\mu\nu}{}^I$ and $L_I$ or $L_I{}^a$, respectively. All of these sectors consistently yield $\xi$ in (3.14). Similarly, $\eta$ in (3.18) is determined in the same way in [19], $i.e.$, the cancellation of the sector $(\overline{\psi}_\mu \gamma^{\mu\nu\rho} T_i \epsilon)(\partial_\nu \phi)(\partial_\rho \phi)$ in the variation of the lagrangian (3.19).

Even though we have relied on the results in the conventional case [10][11][12] for derivation, our system has also some basic differences from the latter. One important difference is the presence of the modified field strength $G_{\mu\nu\rho}$ in (3.20) instead of the explicit Chern-Simons term $C_{IJK} F^I \wedge F^J \wedge A^K$ in the lagrangian in [10][11][12].

As has been already briefly mentioned, there are other fundamental differences of our system from the conventional formulation [10][11][12]. The most important one is the scaling invariance (2.5) of the dilaton $\sigma$ maintained in our (3.19). In the conventional formulation [10][11][12], there is no distinction of the dilaton field $\sigma$ treated separately from other scalar fields, but instead all the scalars form on an equal footing the coordinates of the target manifold specified by the hypersurface $C_{IJK}X^I X^J X^K = 1$ [10][11][12]. In our system (3.19), however, the dilaton keeps its particular role distinguished from other scalar fields. Another important feature is the role of the antisymmetric tensor $B_{\mu\nu}$ which is not mixed up with other vector fields, either. Additionally, all the tensor fields given in [11][12] satisfy the symplectic 'self-duality' condition, always forming pairs with their 'dual' components, while our tensor field $B_{\mu\nu}$ appears as a singlet field.

There is a more technical but more strict comparison of our system with the conventional formulations [10][11][12]. This can be done as follows: We first 'unify' the coordinates $\varphi^\alpha$ with the dilaton $\sigma$ into $\varphi^{\hat\alpha}$ ($\hat\alpha, \hat\beta, \cdots = 1, 2, \cdots, n+1$), following a similar procedure in 9D given by eqs. (3.18) - (3.32) in [17]. Next we perform a duality transformation [18] from $B_{\mu\nu}$ into its dual $B_\mu$, which is now identified with $B_\mu \equiv A_\mu{}^{n+1}$, i.e., $\hat I = n+1$ in the extended set $A_\mu^{\hat I}$ ($\hat I = 1, \cdots, n+1$). At this stage, there are $n+1$ coordinates $\varphi^{\hat\alpha}$ with $n+2$ vector fields $(A_\mu{}^0, A_\mu^{\hat I})$ with an explicit Chern-Simons term $L_{IJ} A^{n+1} \wedge F^I \wedge F^J$ in the lagrangian similar to $C_{IJK} A^I \wedge F^J \wedge F^K$ in [10]. In other words, our $L_{IJ}$ effectively corresponds to $C_{n+1\, IJ}$ [10] in their own notation. Accordingly, the gaugini $\lambda^a$ are unified with $\chi$ to form $\lambda^A$ ($A = 0, 1, \cdots, n$). Consequently, since the numbers of the $\sigma$-model coordinates $\varphi^{\hat\alpha}$, the number of the gaugini $\lambda^A$ and the 'unified' vector fields agree with the general case covered in [10], one might think this is just a special case of the conventional formulation [10][11][12], as if the coset were now enlarged to $SO(n+1,1)/S(n+1)$ or something similar. However, we can easily see why this is not the case, with some fundamental differences. The most typical difference is that the scalar fields $\varphi^{\hat\alpha}$ are no longer the coordinates of the hypersurface of the cone $C$ described in [10][11][12], but they are the coordinates of the *entire* solid cone $C$, as has been also mentioned in a similar system in 9D in [17].[7] Therefore, our system is *not* a special case covered in the conventional formulation [10][11][12].

To be more specific, complying also with the notation given in eqs. (3.18) - (3.32) in [17] as much as possible, we first redefine

$$\widetilde L_I = e^{-\sqrt{3}\sigma} L_I \ , \quad \widetilde L^I = e^{+\sqrt{3}\sigma} L^I \ , \quad \widetilde a_{IJ} \equiv e^{-2\sqrt{3}\sigma}(L_I{}^a L_{Ja} + L_I L_J) \ . \tag{3.23}$$

Here our $L_I$, $L^I$, $L_I{}^a$, $L_a{}^I$ respectively correspond to $h_I$, $h^I$, $h_I{}^a$, $h_a{}^I$ in [17] or [10], and using this rescaled $\widetilde L^I$ ($I = 0, 1, \cdots, n$) is equivalent to the usage of the $(n+1)$-dimensional $\varphi^{\hat\alpha}$-coordinates [10][11][12]. Upon this field redefinition, our $\varphi$ and $\sigma$-kinetic terms in (3.21) are unified as

$$-\tfrac{1}{2} g_{\alpha\beta}(\partial_\mu \varphi^\alpha)(\partial^\mu \varphi^\beta) - \tfrac{3}{4}(\partial_\mu \sigma)^2 \quad \longrightarrow \quad -\tfrac{1}{4} \widetilde a_{IJ}(\partial_\mu \widetilde L^I)(\partial^\mu \widetilde L^J) \ , \tag{3.24}$$

corresponding to (3.20) in [17]. At this stage our system *appears* to be just a special case covered by [10], because all the relevant index-ranges are the same as a special case in [10]. The Chern-Simons term coefficient $L_{IJ}$ in our case satisfies the constraint

$$- L_{IJ} L^I L^J = 1 \ , \tag{3.25}$$

as confirmed by (3.5) - (3.9), as an analog of $C_{n+1\,JK} X^{n+1} X^J X^K = 1$ in [10] or to $C_{IJ} h^I h^J = 1$ in [17]. To have a real comparison with [10], (3.25) is to be rewritten in terms of the whole

---

[7]We acknowledge P.K. Townsend to clarify this point to us.

$\sigma$-model coordinates $\varphi^{\hat{\alpha}}$ or equivalently $\widetilde{L}^I$, as

$$-L_{IJ}\widetilde{L}^I\widetilde{L}^J = e^{2\sqrt{3}\sigma} \quad . \tag{3.26}$$

However, the r.h.s. here is a function of $\sigma$ generally different from unity. Similarly to eq. (3.20) in [17], our (3.26) implies that the unified coordinates $\widetilde{L}^I$, and therefore $\varphi^{\hat{\alpha}}$ are the coordinates of the $(n+1)$-dimensional *entire* solid cone $C$, but not only of the $n$-dimensional surface of the cone $C$ as in [10][11][12]. In other words, one of our coordinates $\sigma$ describes the direction away from the surface of the cone $C$, extending to the *entire* solid cone that has not been covered by the $\sigma$-model coordinates in [10][11][12]. From this result, we can conclude that our system (3.19) is *not* covered as a special case of the formulations in [10][11][12].

This feature of our system is also associated with the above-mentioned special features of the dilaton $\sigma$ and anti-symmetric tensor $B_{\mu\nu}$, which can *not* be unified into a common coordinates with other original $\sigma$-model coordinates or vector fields, as done in [10]. To put it differently, the dilaton and the antisymmetric tensor fields in our system are playing essential roles distinct from other fields in the vector multiplets, as the natural massless NS fields in superstring [1]. ¿From these considerations, we conclude that our system is more natural as the supergravity limit of superstring theory than the conventional formulations [10][11][12].

## 4. Gauging $SO(2)$ Subgroup of $Sp(1)$

We have not yet considered the possible gauging of any subgroup of various gauge groups in our system. In what follows, we consider the gauging of $SO(2)$ subgroup of the automorphism group $Sp(1) = SL(2,\mathbb{R})$, following the similar procedure in [10]. Due to some complications to be explained later, we have to turn off the hypermultiplet from the system, when we consider this $SO(2)$ gauging.

As explained in [10], the $SO(2)$-gauging is performed by introducing the constant vectors $V^I$, with the coupling constant $g$. Accordingly, the covariant derivatives on $Sp(1)$ non-invariant fermions acquire the $SO(2)$ minimal couplings[8]

$$\mathcal{D}_\mu \epsilon^A \equiv D_\mu \epsilon^A + gV_I A_\mu{}^I (T_2\epsilon)^A \quad , \quad \mathcal{D}_{[\mu}\psi_{|\nu]}{}^A \equiv D_{[\mu}\psi_{|\nu]}{}^A + gV_I A_{[\mu|}{}^I (T_2\psi_{|\nu]})^A \quad ,$$
$$\mathcal{D}_\mu\chi^A \equiv D_\mu\chi^A + gV_I A_\mu{}^I (T_2\chi)^A \quad , \quad \mathcal{D}_\mu\lambda^{aA} \equiv D_\mu\lambda^{aA} + gV_I A_\mu{}^I (T_2\lambda^a)^A \quad . \tag{4.1}$$

Here $D_\mu$ is the previous covariant derivatives in (3.21), and the matrix $T_2$ is the second anti-hermitian generator of $Sp(1) = SL(2,\mathbb{R})$ for the $SO(2)$ gauging. The coupling constant $g$ is for the gauging of $SO(2) \subset Sp(1) = SL(2,\mathbb{R})$, while the vectors $V_I$ are all constants, following the similar gauging method in [10][11][12]. Finally, $\xi^{\underline{\alpha}}$ is the Killing vector in the direction of $T_2$ among the generators of $Sp(1)$.

The new explicitly $g$-dependent terms[9] needed in the lagrangian are

$$\mathcal{L}_g \equiv -\tfrac{1}{8}g^2 e^{2\sigma} V_I V_J L^{IJ} - \tfrac{i}{2\sqrt{2}} g e^\sigma (\overline{\psi}_\mu \gamma^{\mu\nu} T_2 \psi_\nu) V_I L^I - \tfrac{1}{\sqrt{2}} g e^\sigma (\overline{\psi}_\mu \gamma^\mu T_2 \lambda^a) V_I L_a{}^I$$

---

[8] Note that we have switched off the hypermultiplets, so that there is no term with $\phi^{\underline{\alpha}}$, etc. in this section.

[9] The word 'explicitly' here means any term with $g$ other than the covariant derivatives (4.1).

$$+ \tfrac{i}{2\sqrt{2}} g e^\sigma (\overline{\lambda}^a T_2 \lambda_a) V_I L^I - \tfrac{2i}{\sqrt{6}} g e^\sigma (\overline{\chi} T_2 \lambda^a) V_I L_a{}^I$$
$$- \tfrac{i}{6\sqrt{2}} g e^\sigma (\overline{\chi} T_2 \chi) V_I L^I + \tfrac{1}{\sqrt{6}} g e^\sigma (\overline{\psi}_\mu \gamma^\mu T_2 \chi) V_I L^I \quad , \tag{4.2}$$

so that the total lagrangian is $\mathcal{L}_0 + \mathcal{L}_g$, now invariant under the transformation rule (3.17) plus the new *explicitly* $g$-dependent terms in the transformation rule of fermions:

$$\delta_Q \psi_\mu |_g = -\tfrac{i}{3\sqrt{2}} g e^\sigma (\gamma_\mu T_2 \epsilon)^A V_I L^I \quad , \qquad \delta_Q \chi^A |_g = +\tfrac{1}{\sqrt{6}} g e^\sigma (T_2 \epsilon)^A V_I L^I \quad ,$$
$$\delta_Q \lambda^{aA} |_g = -\tfrac{1}{\sqrt{2}} g e^\sigma (T_2 \epsilon)^A V_I L^{aI} \quad . \tag{4.3}$$

As usual in gauged supergravity models [9], the gauge coupling $g$ also rescales under (2.5) as $g \to e^{-c} g$, and this explains the function $e^\sigma$ accompanying $g$. One of the crucial relations in the derivations above is

$$[D_\mu, D_\nu] \epsilon_A = -\tfrac{1}{4} R_{\mu\nu}{}^{rs}(\widehat{\omega}) \gamma_{rs} \epsilon_A + g V_I F_{\mu\nu}{}^I (T_2 \epsilon)_A + (D_\mu \phi^{\underline{\alpha}})(D_\nu \phi^{\underline{\beta}}) F_{\underline{\alpha}\underline{\beta}}{}^i (T_i \epsilon)_A \quad , \tag{4.4}$$

which is parallel to the 6D case [19], or more directly to the 5D case [11][12].

Note that our lagrangian (4.2) has the peculiar potential term

$$\mathcal{V}_{\text{pot}} \equiv +\tfrac{1}{8} g^2 e^{2\sigma} V_I V_J L^{IJ} \quad . \tag{4.5}$$

Since $V_I$ are constant, and so is the metric $L^{IJ}$, this $\mathcal{V}_{\text{pot}}$ has the field-dependence only *via* the dilaton $\sigma$. Due to the indefinite metric $(L_{IJ}) = \text{diag.}(-,+,+,\cdots,+)$, the signature of this potential can be flipped, depending on the choice of $V_I \neq 0$. For example, if we choose only $V_0$ to be nonzero, then the potential is negative definite with the AdS background, while if only $V_1, V_2, \cdots, V_n$ are nonzero with $V_0 = 0$ maintained, then $\mathcal{V}_{\text{pot}}$ is positive definite with the dS background. This signature flipping is important for the supersymmetric Randall-Sundrum scenario [5][6][7]. Compared with [10][11][12], our lagrangian is much simpler, depending only on the constant vectors $V_I$ simplifying the couplings to vector multiplets drastically.

We mention here the previously-mentioned complication with the $SO(2)$-gauging in the presence of the hypermultiplets. When the hypermultiplets are included, we encounter $g$- and hypermultiplet-dependent terms that complicate the invariance confirmation. For example, there arises a term with the structure $g \psi^{\underline{a}} D \phi^{\underline{\alpha}}$ out of the variation of the gravitino in the Noether coupling $(\overline{\psi}_\mu{}^A \gamma^\nu \gamma^\mu \psi^{\underline{a}}) V_{\underline{\alpha}\underline{a}A} D_\nu \phi^{\underline{\alpha}}$. This subtlety seems to be also related to the ortho-normality relation mentioned in the footnote before (3.16). The gauging with the hypermultiplets with these subtleties is now in progress.

## 5. Supergravity in Singular Space-Time

As a preliminary for possible supersymmetric Randall-Sundrum scenario [6] we can generalize our system to supergravity in 'singular' space-time, following the prescription in [7]. We start with replacing the original $SO(2)$ gauging coupling constant $g$ *everywhere* by a space-time-dependent scalar field $G(x)$ [7] symbolized as $\mathcal{L}_g \to \mathcal{L}_G$, and then introduce a fourth-rank antisymmetric tensor potential $A_{\mu\nu\rho\sigma}$, with a new term in the lagrangian [7]

$$S_{AG} \equiv \int d^5 x \, \mathcal{L}_{AG} \equiv \int d^5 x \left( \tfrac{1}{24} \epsilon^{\mu\nu\rho\sigma\tau} A_{\mu\nu\rho\sigma} \partial_\tau G \right) \quad . \tag{5.1}$$

The total 5D bulk action $S_{\text{bulk}} \equiv S_0 + S_G + S_{AG} \equiv \int d^5 x \, (\mathcal{L}_0 + \mathcal{L}_G + \mathcal{L}_{AG})$ is no longer invariant under supersymmetry, but has terms proportional to $\partial_\mu G$ [6][7]:[10]

$$\delta_Q(\mathcal{L}_0 + \mathcal{L}_G + \mathcal{L}_{AG}) = \epsilon^{\mu\nu\rho\sigma\tau}\Big[ - i(\overline{\epsilon}\gamma_{\mu\nu}T_2\psi_\rho)V_I A_\sigma{}^I + \tfrac{1}{6\sqrt{2}} e^\sigma (\overline{\epsilon}\gamma_{\mu\nu\rho}T_2\psi_\sigma)V_I L^I$$
$$+ \tfrac{i}{24\sqrt{6}} e^\sigma (\overline{\epsilon}\gamma_{\mu\nu\rho\sigma}T_2\chi)V_I L^I - \tfrac{i}{24\sqrt{2}} e^\sigma (\overline{\epsilon}\gamma_{\mu\nu\rho\sigma}T_2\lambda^a)V_I L_a{}^I \Big]\partial_\tau G \,. \quad (5.2)$$

This is to be cancelled by the new supersymmetry transformation rule $\delta_Q A_{\mu\nu\rho\sigma}$ in $\mathcal{L}_{AG}$:

$$\delta_Q A_{\mu\nu\rho\sigma} = + 24 i(\overline{\epsilon}\gamma_{[\mu\nu|}T_2\psi_{|\rho|})V_I A_{|\sigma]}{}^I - 2\sqrt{2} e^\sigma (\overline{\epsilon}\gamma_{[\mu\nu\rho|}T_2\psi_{|\sigma]})V_I L^I$$
$$- \tfrac{i}{\sqrt{6}} e^\sigma (\overline{\epsilon}\gamma_{\mu\nu\rho\sigma}T_2\chi)V_I L^I + \tfrac{i}{\sqrt{2}} e^\sigma (\overline{\epsilon}\gamma_{\mu\nu\rho\sigma}T_2\lambda^a)V_I L_a{}^I \,, \quad (5.3)$$

while we maintain $\delta_Q G = 0$. Due to the additional field $\chi$ in our system compared with [7], we have four terms in total in (5.3). After this, the action $S_{\text{bulk}}$ is adjusted to be superinvariant.

In order to generalize this result to more singular 5D space-time like Randall-Sundrum solution [5], we now add the brane action $S_{\text{brane}}$ to $S_{\text{bulk}} \equiv S_0 + S_G + S_{AG}$, which characterizes the space-time singularities:

$$S_{\text{brane}} = -2\widetilde{g} \int d^5 x \left[ \delta(x^5) - \delta(x^5 - b) \right] \left( a e_{(4)} V_I L^I + \tfrac{1}{24} \epsilon^{\mu\nu\rho\sigma\,5} A_{\mu\nu\rho\sigma} \right) \,, \quad (5.4)$$

where $\widetilde{g}$, $a$ and $b$ are constants, and $e_{(4)}$ is determinant of the 4D vierbein embedded in the fünfbein $e_\mu{}^m$. This modifies the original field equation of $A_{\mu\nu\rho\sigma}$ from $\partial_\mu G = 0$ into

$$\partial_5 G(x^5) = 2\widetilde{g} \left[ \delta(x^5) - \delta(x^5 - b) \right] \,, \quad (5.5)$$

with the solution [6][7]

$$G(x) = \widetilde{g}\, \epsilon(x^5) = \begin{cases} +\widetilde{g} & (\text{for } 0 < x^5 \leq +b) \,, \\ -\widetilde{g} & (\text{for } -b \leq x^5 < 0) \,. \end{cases} \quad (5.6)$$

As for the explicit solutions for Killing spinor equations consistent with the Randall-Sundrum brane solution [5], we have the same situation as the conventional case [6][7][8].[11] Namely, we have subtlety about consistent solutions satisfying both the gravitational equation and the Killing spinor equations simultaneously [6][7][8], associated with the integrations involving the signature function $\epsilon(x^5)$. Therefore we do not elaborate this aspect of our theory any further.

## 6. Concluding Remarks

In this paper, we have presented an alternative $N = 2$ supergravity multiplet with $12 + 12$ degrees of freedom, coupled to $n$ copies of vector multiplets in 5D, and hypermultiplets, with a simpler coupling structure compared with the conventional supergravity

---

[10] When the $SO(2)$-gauging is present, we truncate the hypermultiplets, as mentioned before.

[11] We acknowledge J. Bagger and M. Zucker clarifying this situation.

[10][11][12]. These couplings are parallel to the 9D case [16], in which the scalars in the vector multiplets form the coordinates of the $\sigma$-model for the non-Jordan family scalar coset $H^n \equiv SO(n,1)/SO(n)$, and the vector fields with the total number $n+1$ form the $(\mathbf{n+1})$-representation of $SO(n,1)$, while the gaugini $\lambda^a$ form the $\mathbf{n}$-representation of $SO(n)$. The scalars in the hypermultiplets form the $\sigma$-model on the quaternionic Kähler manifold $Sp(n',1)/Sp(n') \times Sp(1)$. We have also performed the gauging of the $SO(2)$ subgroup of $Sp(1) = SL(2,\mathbb{R})$ in the absence of the hypermultiplets, with a peculiar potential term in the lagrangian. We have also generalized this result to the case of singular space-time as a preliminary for supersymmetric Randall-Sundrum scenario [5] similarly to the conventional 5D supergravity [6][7][8].

Compared with the conventional formulation of $N=2$ supergravity in 5D [10][11][12], we can summarize here the several differences in our formulation:

i) There is *no* explicit Chern-Simons term such as $C_{IJK} F^I \wedge F^J \wedge A^K$ in our lagrangian in contrast to the conventional case [10][11][12]. Instead, the Chern-Simons term is implicit in the field strength $G_{\mu\nu\rho}$.

ii) Our scalar potential for the gauged case is much simpler than that in [10][11][12].

iii) The couplings of our antisymmetric tensor $B_{\mu\nu}$ and dilaton $\sigma$ fields are much more like those in superstring theory [1], with the manifest global scaling property (2.5).

iv) In our formulation, the vector and tensor multiplets are *not* unified on an equal footing as in [10][11][12].

v) It seems that our system is not equivalent to the conventional one [10][11][12], covered as a special case.

Some remarks are now in order: As for the point i), the Chern-Simons form appears only in our field strength $G_{\mu\nu\rho}$ but not explicitly in the lagrangian. Our system is much like $N=1$ supergravity [16] in 9D, in which $L_{IJ}$ also controlled the system. As for the point ii), our potential is simpler, because it depends only on the dilaton $\sigma$, because of the constancy of the 'metric' $L_{IJ}$. There is no further complication by the scalar field, in contrast to [10][11][12] with the non-trivial coefficients $C_{IJK}$. As for the point iii), this is the advantage of our formulation with the usual antisymmetric tensor and dilaton couplings with scaling properties, as is naturally expected from the usual superstring theory [1]. As for the point iv), we point out that the vector and tensor multiplets are more or less on an equal footing in [11][12], *e.g.*, the spin 1/2 fields are forming the $(\mathbf{n}_V + \mathbf{n}_T)$-representation of $SO(n_V + n_T)$ for $n_V$ and $n_T$ copies of vector and tensor multiplets. The difference in scaling weights between the vectors and tensors in our system also forbids such an unified treatment of these multiplets. The point v) is also supported by many important facts. First, the scaling weight $(+2)$ for $B_{\mu\nu}$ as seen from (2.5), and therefore $(-2)$ for its Hodge dual $B_\mu$ [18] is different from the scaling weight $(+1)$ for other vector fields $A_\mu{}^I$, forbidding a unilateral treatment of $B_\mu$ unified with other vectors. Second, as was also mentioned, the comparison with the conventional formulation [10] can be more easily done, by first unifying the dilaton $\sigma$-field with other coordinates $\varphi^\alpha$ into $\varphi^{\hat\alpha}$ following the similar procedure in [17], and next performing the duality transformation [18] of $B_{\mu\nu}$ into $B_\mu$, ending up with a Chern-Simons term similar to that in [10], following eqs. (3.18) - (3.32) in [17]. Even though this result looks just like a special case covered in [10] *superficially* with the same field content, there is a fundamental difference about the manifold. Because the $\sigma$-model coordinates thus obtained will be no longer those for the hypersurface of the cone $C$, but they are for the *entire* cone, which has *not* been covered as a special case in [10][11][12], as explained more rigorously in section 3. Third, the various exponential factors of the dilaton $\sigma$ (also related to the point iv) above) does not seem to be absorbed into the redefinitions of the geometric quantities of the coset. All of these are traced back to the dilaton and the antisymmetric tensor playing essential roles as peculiar NS massless fields, distinguished from other fields in the vector multiplets, indicating that our system is 'closer' to superstring theory [1].

Some readers may wonder why our peculiar supergravity multiplet has not been so far covered as a special case in the conventional formulation [10][11][12] which has been so exhaustively studied. This is, however, understandable from the viewpoint that the original work in [10] was initiated before the discovery of phenomenological importance of superstring in 1984 [24][1]. Therefore, there was no strong motivation to include the dilaton and antisymmetric tensor fields with particular importance in the system. In other words, it is only superstring [1] or M-theory [2] that motivates the peculiar couplings of dilaton and antisymmetric tensor to supergravity, as we have performed in the present paper.

Even though we have stressed the difference of our formulation from other general matter couplings in [10][11][12], it is fair to mention also some similarity. For example, we expect it possible to generalize the number of the additional tensor fields $B_{\mu\nu}$ in addition to the one in the supergravity multiplet with a $\sigma$-model structure similar to that presented in [11][12].[12] In other words, not only for the vector field $A_\mu$ in our $12+12$ irreducible supergravity multiplet, but also for the tensor field $B_{\mu\nu}$, we can couple some copies of outside 'matter' multiplets, and get some non-compact $\sigma$-model structure. The generalizations in these directions are now under way.

We are grateful to S.J. Gates, Jr. for helpful discussions. We are indebted to A. Ceresole, G. Dall'Agata, M. Günaydin and P.K. Townsend for clarifying matter multiplet couplings, while to J. Bagger and M. Zucker for informing about the supersymmetric Randall-Sundrum scenario. Special acknowledgement is for M. Luty for bringing about the subject of 5D supergravity, and also for E. Sezgin for his important suggestions. Additionally, we acknowledge the referee of this paper who devoted his time to the improvement of its content.

---

[12]To avoid misinterpretation, we stress that our antisymmetric tensor $B_{\mu\nu}$ is still distinguished from these additional ones, which always appear in pairs [11][12].